\documentclass[aps,preprint,amssymb,12pt,floatfix]{revtex4}
\usepackage{subfigure}
\usepackage{graphicx}
\usepackage{rotating}
\usepackage{color}
\usepackage{amsmath, amsthm}
\usepackage{epstopdf}
\begin{document}

\title{Crowding effects on the structural transitions in a flexible helical homopolymer}
\author{Alexander Kudlay$^1$, Margaret S. Cheung$^2$ and D. Thirumalai$^{1,3}$}
\affiliation{$^1$Biophysics Program, Institute for Physical Science and Technology, University of Maryland, College Park, Maryland 20742\\
$^2$ Physics Department, University of Houston, Houston, TX 77204\\
$^3$ Department of Chemistry and Biochemistry, University of Maryland, College Park, MD 20742}

\date{\small \today}

\baselineskip = 21pt
\begin{abstract}
We elucidate the structural transitions in a helical off-lattice homopolymer induced
by crowding agents, as function of the number of monomers  ($N$) and volume fraction ($\phi_c$)
of crowding particles.   At $\phi_c = 0$,  the homopolymer
undergoes transitions from a random coil to a helix, helical hairpin \textbf{HH}, and helix bundle \textbf{HB} structures depending on $N$, and temperature. Crowding induces chain compaction that can promote \textbf{HH} or \textbf{HB} formation
depending on $\phi_c$.  Typically,  the helical content decreases which is reflected
in the decrease in the transition temperatures that depend on  $\phi_c$, $N$, and the size of the crowding particles. 

\baselineskip 21pt
\end{abstract}
\maketitle
\newpage

The volume fraction ($\phi_c$) of large macromolecules
such as lipids, ribosome,  and  cytoskeleton
fibers \cite{Baumeister02} in the cell interior, which  can be as large as 0.4 \cite{Ellis03}, affects all
biological processes ranging from transcription to folding of RNA and proteins. Protein stability \cite{Minton00COSB,Zhou04} and folding rates \cite{Cheung05} of proteins are
enhanced by an entropic stabilization mechanism (ESM) according to which the predominant contribution to
the native state stabilization is due to an increase in the free energy of the unfolded states.  Entropy decrease  of the unfolded states results
from the suppression of the number of allowed conformations of the polypeptide chains due to volume excluded by
the crowding particles, while the native state is affected to a lesser extent.  
The ESM \cite{Minton00COSB,Cheung05} is  linked to crowding agent-induced depletion attraction \cite{Oosawa54,Cheung06JMB,Kamien05, Cheung05}
between the  monomers of the protein or RNA \cite{Pincus07}.   Crowding
agents can also profoundly affect protein-protein interactions \cite{Rivas01Kozer0407} and amyloid
formation \cite{Hatters02} that is linked to a number of  neurodegenerative diseases. 

We consider crowding
effects on one the random coil \textbf{RC} to helix \textbf{H} transition. 
The interplay between the multitude of interactions 
between the crowding agents and
proteins ($V_{CP}$) and the intra-protein forces ($V_{P}$) makes it
difficult the structural changes that occur in a protein when
$\phi _{c}\neq 0$. 
We consider the effect of spherical crowding agents on an off-lattice model of a homopolymer chain \cite{Guo96Klimov98},  which
undergoes a coil to helix transition as temperature ($T$) is varied when $\phi
_{c}=0$. 
The major results, which were obtained
for polymers with different $N$ and $\phi_c$ using molecular simulations,  are:

(a) The phase diagram is determined by a balance between the
strength, $\gamma$,  of the dihedral angle potential that is related to the local stiffness and the parameter $\delta $, which
specifies the strength of the hydrophobic attraction between the non-bonded beads, i.e. ones
that are separated by three or more covalent bonds. 
As $\delta $ is varied the
homopolymer undergoes a series of structural transitions  from  a \textbf{RC} to \textbf{H},
helical hairpin (\textbf{HH}), and helix bundle (\textbf{HB}) at low temperatures, depending on $N$. For a fixed $\phi_c$, crowding
particles whose  radius ($r_c$) is commensurate with the the size of the monomer($r_m$) \cite{Oosawa54, Oosawa58}  ($\frac{r_m}{r_c} \approx 1$),   have the largest effect in stabilizing the collapsed structures. 


(b) The $r_c$ and $\phi_c$-dependent transition temperature, $T_{S}(r_c, \phi_c)$, from random coil (\textbf{RC}) to predominantly helical
conformations (\textbf{HH} or \textbf{HB}) changes dramatically depending on $%
r_{c}$ and $\phi _{c}$. For a fixed $\phi _{c}$, the most significant change
in $T_{S}(r_c, \phi_c)$ (compared to $\phi _{c}=0$) occurs when $r_{c}$ is on the order of the
size of the monomer. The values of $T_S(r_c, \phi_c)$ saturate, at all $N$, when $r_c$ becomes large. At a fixed $r_{c}$,  $T_{S}(r_c, \phi_c)$ decreases
as $\phi _{c}$ increases.

The conformations of a homopolymer chain, with
$N$ connected beads \cite{Guo96Klimov98}, 
The chain conformations are specified by the vector $\{r_{i}\}$ , \ $%
i=1,2,..N$. The potential energy of the chain is a sum of bond-stretch potential, bond-angle
potential, interactions associated with the ($N-3$) \ dihedral angle degrees
of freedom ($V_{D}$), and non-bonded potential ($V_{N}$) that determines the
extent of tertiary interactions. The energy functions  $V_D$ and $V_N$ are %
$V_{D}=\sum_{i} \gamma \varepsilon \left\{ (1+\cos \left( \phi_i +\frac{%
2\pi }{3}\right) )+(1+\cos 3\phi_i )\right\}$
and  $V_{N}=\sum_{i\neq j}\varepsilon \left[ \left( \frac{2r_{m}}{r_{ij}}%
\right) ^{12}-2\delta \left( \frac{2r_{m}}{r_{ij}}\right) ^{6}\right]$
where $\varepsilon $ (=1kcal/mol) specifies the energy of interactions\ between
non-bonded beads $i$ and $j$ separated by $r_{ij}=|r_{i}-r_{j}|$, $\gamma$ (=1 in this work) is
the strength of the of the dihedral potential, $\phi_i$ is the $i^{th}$ dihedral angle, and  
$r_{m}=2$ \AA\ is the size of a monomer.  The potential $V(=V_{CC}+V_{CP}$),
arising from interactions between the spherical crowding
particles ($V_{CC}$) and with the monomers ($V_{CP}$) is
$V=\sum_{i\neq j}\varepsilon \left[ \left( \frac{2r_{c}}{r_{ij}}%
\right) ^{12}\right]+\sum_{i, j}\varepsilon \left[ \left( \frac{r_{c}+r_{m}}{r_{ij}^{CP}}\right) ^{12}\right]$
\noindent where $r_{ij}$ is the distance between the crowding particles $i$ and $j$, $r_{ij}^{CP}$ is the distance between bead $i$ and the crowding particle $j$, and $r_c$ is size of the crowder.

The simulations were performed using a modified in-house AMBER6 \cite{AMBER} package that was altered to
incorporate Langevin dynamics in the low friction limit \cite{Klimov97} to enhance the rate of conformational sampling \cite{Veishans97}. 
 In the presence
of crowders the calculations were performed in the NVT ensemble. The homopolymer and the crowding
particles are confined to a cubic box and periodic boundary conditions are
used to minimize surface effects. The size of the simulation box, is determined by the condition that the
box contain a minimum of 150 crowding particles. For small $r_{c}$, the edge of
the box is equal to the sum of the length of the fully extended helix and
four times the average distance between the crowders at a specified $\phi
_{c}$. The number of crowding particles ranges from $\approx 150$ for the most dilute system with
the largest $r_c$, to 1200 ($\phi_c =0.2$, $r_{c}=2$ $\mathring{A}$, $%
N=16$), and  3000 ($\phi_c =0.2$, $r_{c}=4$ $\mathring{A}$,
$N=64$). 

The sampling efficiency in the simulations are enhanced using the replica exchange method (REM)\cite{Cheung05,Sugita99}, which ensures that the thermodynamic averages are fully converged.  We used twenty replicas in the temperature range from T=100 to 400K in the REM simulations. The initial configurations are randomly chosen from high-temperature simulations, and subsequently quenched to the desired temperatures. The integration time step is $10^{- 4}  \tau_L$ where $\tau_L = (m r_m^2 /\varepsilon)^{\frac{1}{2}}$ with m being  the mass of a bead. At chosen time interval (= $40 \tau_L$), configurations with neighboring temperatures are exchanged. The acceptance probability, which depends on the temperatures and the energies of the replicas, was in the range $0.2 - 0.3$.  In order to calculate averages we retained between (4,000 - 8,000) conformations for each replica. The results of the REM simulations were combined with independent data set generated using the weighted histogram analysis method (WHAM)\cite{WHAM} to obtain thermodynamic averages. The structural transitions in the hompolymer are characterized by using the
specific heat $C_{V}$ and the radius of gyration $R_{g}$.  The extent 
of helical order was quantified using 
$f_{H}(T,\phi _{c})=\frac{1}{N-3}\sum_{i=1}^{N-3}\left\langle \Theta (\Delta \phi -|\phi _{i}-\phi _{i}^{N}|)\right\rangle$
where $\Theta (x)$ is the Heavyside function,  $\phi _{i}^{N}$ is the
value of the $i^{th}$ dihedral angle in the energy
minimized ($T=0$) helical state, and $\Delta \phi$ is the tolerance
in $\phi_i$ used to assign helical character to the $i^{th}$ dihedral angle. 
We chose $\Delta \phi =12.07%
{{}^\circ}%
$ to ensure that $T_{S}$ obtained using the criterion $%
f_{H}(T_{S}, \phi_c = 0)=0.5$ is consistent with the temperature at which $C_{V}$ for $N=64$ has a maximum. 

The structural transitions as a function of $T$ and $%
\delta $ for $N=16$ with $\phi _{c}=0$ (Fig. 1a) show that the \textbf{RC}$%
\rightarrow $\textbf{H} transition occurs at $T_{S}\approx 292$
K at $\delta =0$. For low to moderate $\delta$ values ($\delta \lesssim 0.5$) the polymer exists either as a \textbf{RC} or a \textbf{H}. With $%
\delta =0.75$, we find a transition to a \textbf{HH} that is
accompanied by a drastic reduction in $R_{g}$. At high $\delta$ values,  the energy cost to form a bend in \textbf{HH} is
compensated by a number of favorable tertiary contacts that stabilize the
\textbf{HH}.  The chain compaction at high $\delta$ 
results in structures that have high helical content as measured.
For $N=32$ and $64$,  besides \textbf{H} and \textbf{HH}s
we find that helix bundles (\textbf{HB}s) can also form as $\delta $ and $T$ are
changed. For some range  of $\delta$ the 
ordered structures coexist, while for  other choices the probability distribution is
peaked around only one unique structure. 
The transition temperature $T_{S}$ at
which the ordered structures form changes dramatically as $\delta $ increases (Fig. 1b).
Only when $\delta >0.5$, do we find significant dependence of $%
T_{S}(\delta ,\phi _{c})$ on $N$ (Fig. 1b). 

From the temperature dependence of the thermally averaged $R_{g}$ with $\phi
_{c}=0.2$ and at various sizes of $r_{c}$ we find that smaller crowding agents ($%
r_{c}=2$ or $4$ \AA ) are most efficient in inducing chain compaction (Fig. 2a). For $r_{c}=2$ \AA\ \ the values of $R_{g}$ even at high
temperatures are considerably smaller than $R_{g}^{H}$ - the radius
of gyration of the energy-minimized helical structure ($T = 0$). A qualitative
explanation follows from  the Asakura-Oosawa (AO) theory \cite{Oosawa54}, which predicts that 
 the strength of the additional entropically-induced effective
attraction between the beads increases as $r_{c}$ decreases and $\phi _{c}$
increases. As a result, the effective attraction $\delta _{R}(\phi
_{c},r_{c})\sim \delta _{0}+f(\phi _{c},r_{c})$, which increases with decreasing $r_c$ \cite{Shaw91}, is largest for small $r_c$.
In the helical homopolymer model a reduction in $R_{g}$, at  sufficiently large $\delta$, is also
accompanied by enhancement in helical order which explains the emergence of \textbf{HH} (Fig. 2b). Thus, crowding agents with 
$\frac{r_m}{r_c} \approx 1$ are most efficient in inducing ordered structure formation at high $T$ even if $\delta$ is not large.

The probability density, $P(T,R_{g}/R_{g}^{H})$, in Fig. 2b for $\phi
_{c}=0.2$, shows that the distribution function changes continuously as $T$
decreases. At high temperatures, there is one broad peak that represents an
ensemble of mostly random coil \textbf{RC} structures. As $T$ decreases below $%
T_{S}(r_c, \phi _{c})~$a sharp peak that corresponds to 
\textbf{HH} structures, with high $f_H(T, \phi_c)$,  emerges. Since there is only a continuous shift
in $P(T,R_{g}/R_{g}^{H})$, without a discernible region of bimodality, the transitions to structures with high
$f_H(T, \phi_c)$ are not  "phase transitions".
Rather, the energy landscape has multiple basins of
attraction with varying helical content whose population can be altered by changing $T$, $\phi _{c}$, or
$\delta $.

Transitions to higher order (three or more) \textbf{HB} structures occur for
$N=64$ with $\phi _{c}=0.2$ and $r_{c}=4$ \AA\ (Fig. 3).  In this case,
crowding-induced formation of \textbf{HH} and \textbf{HB} at low $T$ is also
accompanied by a dramatic reduction in $R_g$ (Fig. 3a). As temperature
decreases, there is evidence for coexistence between long \textbf{HH}
and \textbf{HB} (Fig. 3b). The formation of a large number of inter-helical
contacts compensates for the energetic cost due to bend formation which results in the transition to the \textbf{HB}. 

The volume fraction $\phi _{c}$ can be independently altered by changing
either $r_{c}$ or  the number density of the crowding agents. For
a fixed $N$ and $r_{c}$, the values of $T_{S}(r_c, \phi_c)$
decrease as $\phi _{c}$ increases (Fig. 4a). The variations are larger for
the smaller $r_{c}$ (Fig. 4a). The decrease in $T_{S}(r_c, \phi_c)$
with increasing $\phi _{c}$ is a consequence of the enhancement in $\delta
_{R}(\phi _{c},r_{c})$ caused by the entropic depletion attraction. From the
AO theory, it follows that the strength of the depletion attraction $f(\phi _{c},r_{c})$ should increase
as $\phi_{c}$ increases.  Thus, $T_{S}(r_c \phi_c)$
should have the largest shift as $\phi_{c}$ increases, which is in accord with
our simulations (Fig. 4a).

The
transition temperatures $T_{S}(r_c, \phi_c)$ ( obtained  using  \ $f_H(T_S, \phi_c) = 0.5$)
reports on the total helical content independent of whether the structure is a \textbf{H}, \textbf{HH} or \textbf{HB}. 
The changes in $T_{S}(r_c,\phi_c)$ for a fixed $\phi _{c}$ and varying $%
r_{c}$ are shown in Figure 4A. We expect that as $r_{c}$ increases beyond $%
R_{g}^{H}$ the transition temperature $T_{S}(r_c,\phi_c)$ should approach the value expected for
folding in narrow confined space formed by large crowding agents, and hence
be independent of $r_{c}$. This is borne out by the simulations which show
that $T_{S}(\phi _{c},r_{c})$ is almost constant as $r_{c}>30$ \AA\ (Fig.
4b). For smaller values of $r_{c}$ Fig. 4b shows that $T_{S}$ decreases sharply especially for $N=16$. As in Fig. 4a, we
find that the largest changes are obtained for $r_{c}=2$\AA . 

The stability of helical conformations is determined by interplay of the local stiffness and the
specific attractive interactions between beads $i$ and $i + 3$.  In contrast, crowding agents, which enhance non-specific
homogeneous attraction between the beads, induce chain compaction.  Whether chain compaction is also
accompanied by enhanced helical stability depends on the range and the strength of the AO attraction.  At all values of
$r_c$, the transition temperatures $T_S(r_c, \phi_c)$ decrease as $r_c$ decreases with the change being most dramatic for small $r_c$ (Fig. 4b).  These
results show that the helical stability decreases when $\phi_c$ is non-zero even though the chain is compact.  Thus, we conclude that
the homogeneous AO attraction compromises the forces required to stabilize particular (\textbf{H}, \textbf{HH}, or \textbf{HB}) helical states.  If the polymer backbone were stiff ($\gamma > 1$) then stretches of helical
conformation on the scale of the persistence length of the chain would persist especially at low temperatures \cite{Kamien05}. For the flexible helical polymer there is a loss in helical stability, which is manifested by a decrease in $T_S(r_c,\phi_c)$ that is most pronounced when $r_c$ is small.


We conclude with the following remarks. (1) 
Variation of the  non-zero  hydrophobicity parameter  $\delta $ leads to a variety of higher order
structures at low temperatures, such as \textbf{HH} and \textbf{HB},  whose stabilities can be  enhanced by macromolecular  crowding.
(2) The high temperature denatured structures  are more  compact at $\phi_c \neq 0$ (see Figs. 2a and 3a) than when $\phi_c = 0$, which has implications for crowding-induced folding mechanisms of proteins.  (3) 
The prediction that the helix stability changes, as $\phi_c$ and $r_c$ are varied, can be validated using circular dichroism (CD) spectroscopy that detects the extent of helix formation in proteins \cite{Perham07FEBS}. 

\textbf{Acknowledgements:} We are grateful to E. O'brien  and N. Toan for useful discussions. This work was supported
by a grant from the National Science Foundation through grant number CHE 05-14056. 
M.S.C. appreciates support from the University of Houston.

\newpage
{\LARGE Figure Captions}

\textbf{Figure 1.} (a) Normalized radius of gyration
$R_{g}/R_{g}^{H}$
as a function of temperature for several values of the short-range attraction parameter $%
\delta $ for the $N=16$ chain. Snapshots of typical conformations
are also shown.  The value of $R_{g}^{H} = 10.1$ \AA\ .  (b)  Transition temperature $T_S$ as a
function of  $\delta $ for $N=16,32$ and $64$.

\textbf{Figure 2.} (a) Normalized radius of gyration
$R_{g}/R_{g}^{H}$ as a function of $T$ for different crowder radii
$r_{c}$ (indicated in the Figure) at 
 $\phi_c =0.2$ and $N=16$. (b) Probability distribution functions
of $R_{g}/R_{g}^{H}$ for different temperatures for  $\phi_c =0.2$ and
$r_{c}=2$ \AA\ case in (a).

\textbf{Figure 3.} Same as Fig.2 except N=64. (a)
Normalized radius of gyration $R_{g}/R_{g}^{H}$ (with $R_{g}^{H} = 40.4$ \AA\ ) at constant
$\phi_c=0.2$. (b) Probability distribution functions of
$R_{g}/R_{g}^{H}$ for the $\phi_c =0.2$ and $r_{c}=4$ \AA\ case shown in
(a). 

\textbf{Figure 4.} (a) Helix-coil transition temperature $T_S(r_c,\phi_c)$ as a function of $\phi_c $ at a constant $r_{c}$ for three 
chains with $N=16$
($r_{c}=2\mathring{A}$), $N=32$ ($r_{c}=4\mathring{A}$), and $N=64$ ($r_{c}=4%
\mathring{A}$). (b) $%
T_S(r_c,\phi_c)$ as
a function of the crowder radius $r_{c}$ at 
 $\phi_c=0.2\, $. Plots for chains of lengths $N=16,32$ and $64$ are shown. In both (a) and (b) $\delta = 0.3$.

\newpage
\includegraphics[scale=0.7]{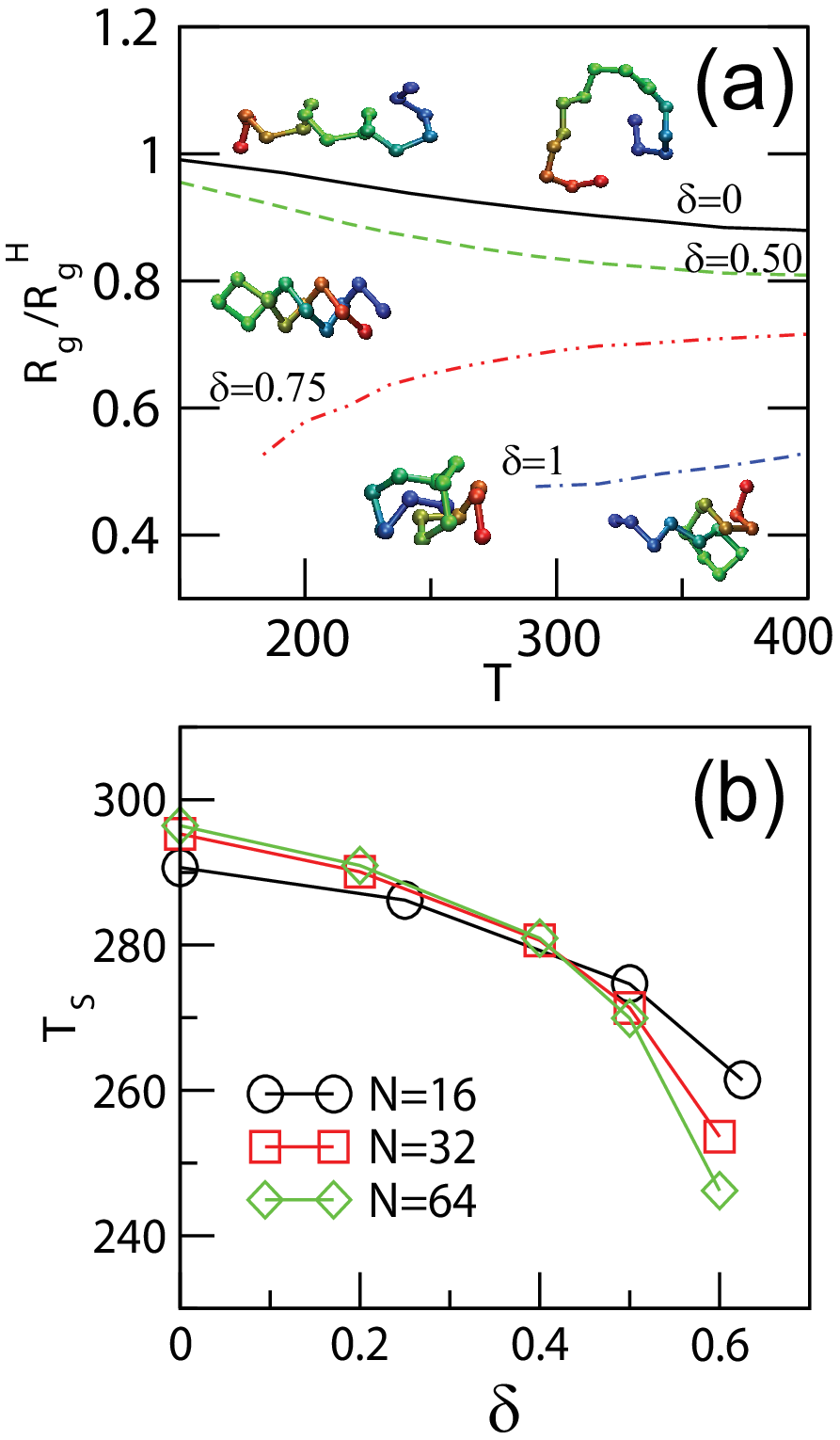} 
\newpage
\includegraphics[scale=0.7]{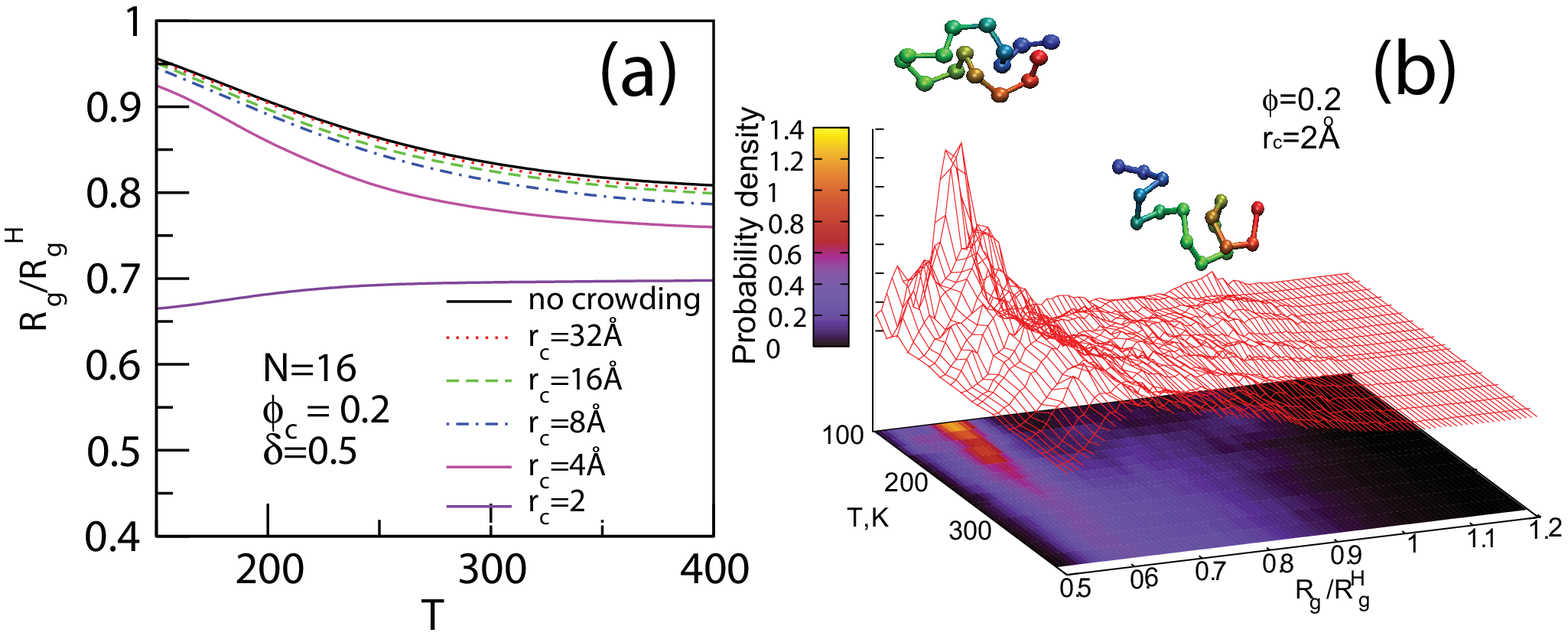} 
\newpage
\includegraphics[scale=0.7]{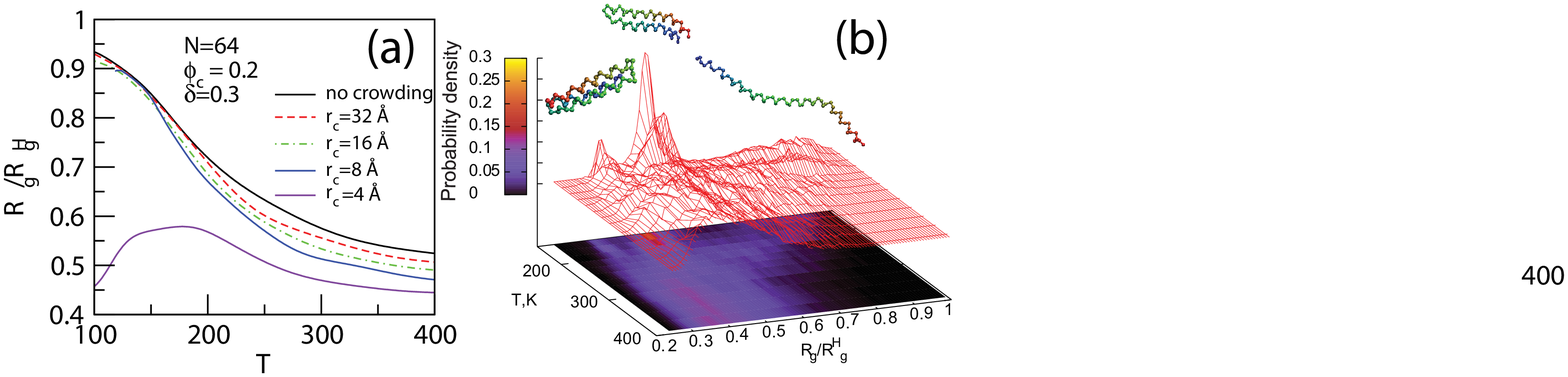} 
\newpage
\includegraphics[scale=0.7]{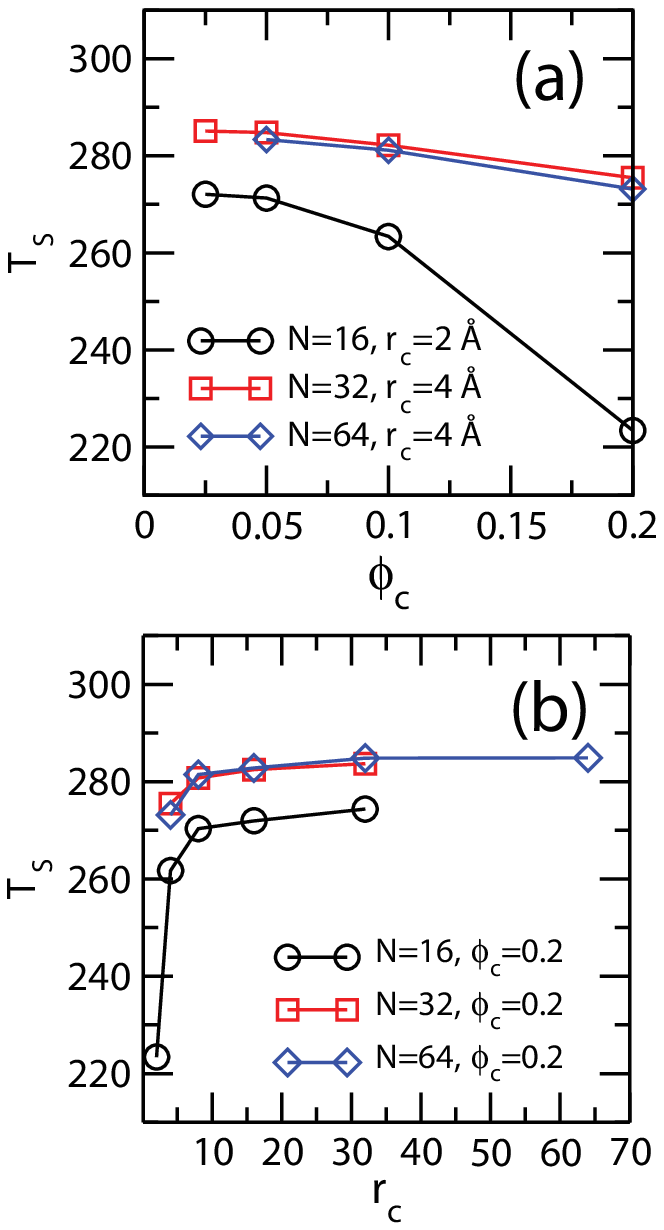} 


\begin{thebibliography}{99}
\bibitem{Baumeister02} O. Medalia,  I. Weber, A. S. Frangakis, D.  Nicastro, and W.  Baumeister,  Science  \textbf{298}, 1209 (2002).

\bibitem{Ellis03} R. J. Ellis and A. P.  Minton, Nature \textbf{425}, 27 (2003).

\bibitem{Minton00COSB} A. P. Minton, Curr. Opin. Struct. Biol. \textbf{11}, 114 (2000).

\bibitem{Zhou04} H. X. Zhou, Acc. Chem. Res.,  \textbf{37}, 123 (2004).

\bibitem{Cheung05} M. S. Cheung,  D. K.  Klimov,  and D.  Thirumalai,  Proc. Natl.
Acad. Sci. U.S.A. \textbf{102}, 4753 (2005).

\bibitem{Cheung06JMB} M. S. Cheung and D. Thirumalai, J. Mol. Biol. \textbf{357}, 632  (2006). 

\bibitem{Oosawa54} S. Asakura and F.  Oosawa, J. Chem. Phys., \textbf{22}, 1255 (1954).

\bibitem{Kamien05} Y. Snir and R. D.  Kamien,  Science \textbf{307}, 1067 (2005).

\bibitem{Pincus07}  D. L. Pincus,  C. Hyeon, and D.  Thirumalai,  J. Amer. Chem. Soc. \textbf{130}, 7364 (2008). 

\bibitem{Rivas01Kozer0407} R. Rivas,  J. A. Fernandez,  A. P.  Minton,  Proc. Natl.
Acad. Sci. U.S.A.   \textbf{98}, 3150. (2001);  N. Kozer and   G.  Schreiber,  J. Mol. Biol.  \textbf{336}, 763 (2004);
N. Kozer, Y. Y.  Kuttner, G.  Haran, and G.  Schreiber,  Biophys. J.  \textbf{92}, 2139 (2007).

\bibitem{Hatters02} D. M. Hatters, A. P.  Minton,  G. J.  Howlett, J. Biol. Chem. \textbf{277}, 7824 (2002).


\bibitem{Minton05} A. P. Minton,  Biophys. J.  \textbf{88}, 971 (2005). 

\bibitem{Sasahara03} K. Sasahara,  P. McPhie and A. P.  Minton,  J. Mol. Biol.  \textbf{326}, 1227 (2003).


\bibitem{Guo96Klimov98} Z. Guo and D.  Thirumalai,  J. Mol. Biol. %
 \textbf{263}, 323 (1996).;  D. K. Klimov,  M. R.  Betancourt and  D. Thirumalai %
Folding \& Design  \textbf{3}, 481 (1998).

\bibitem{Sugita99} Y. Sugita and Y. Okamoto, Chem. Phys. Lett. \textbf{314}, 141 (1999).


\bibitem{Oosawa58} S. Asakura and F.  Oosawa,  J. Polymer Sci.  \textbf{33}, 183 (1958).

\bibitem{AMBER} D. A. Case,  D. A.  Pearlman, J. W.  Caldwell,  T. E.  Cheatham, 
W. S.  Ross,  C.  L. Simmerling,  T. A.  Darden,  K. M.  Merz, R. V. Stanton, 
A. L.  Cheng,  et. al. \textit{AMBER6}, University of California, San
Fransisco, (1999).


\bibitem{Klimov97} D. K. Klimov and D.  Thirumalai,  Phys. Rev. Lett.  \textbf{79}, 317 (1997).

\bibitem{Veishans97} T. Veishans, D. K.  Klimov and  D. Thiruamalai, Fold. Des. \textbf{2}, 1 (1997).


\bibitem{WHAM} S. Kumar, D.  Bouzida, R.  Swendsen,  P. A.  Kollman, 
J. M. Rosenberg,  J. Comput. Chem. \textbf{13}, 1011. (1992); J. D.  Chodera, W. C.  Swope,  J. W.  Pitera, C. Seok,
K. A. Dill,  J. Chem. Theory Comput.  \textbf{3}, 26 (2007).

\bibitem{Shaw91} M. R. Shaw and D.  Thirumalai, Phys. Rev. A  \textbf{44}, R4797 (1991).

\bibitem{Perham07FEBS} M. Perham and P. Wittung-Stafhede, FEBS Lett. \textbf{581}, 5065 (2007).




\end{thebibliography}
\end{document}